# Borges Dilemma, Fundamental Laws, and Systems Biology

Ping Ao
Department of Mechanical Engineering and Department of Physics
University of Washington, Seattle, WA 98195, USA
June 7, 2007; revised August 6, 2007

I reason here that the known folk law in biology that there is no general law in biology because of exceptions is false. The (quantitative) systems biology offers the potential to solve the Borges Dilemma, by transcending it. There have already a plenty of indications on this trend.

A well-known folk-law in biology is that there is no general law in biology because of exceptions. In her recent elegant essay E.F. Keller gave a particular nice presentation on the exceptions to rules or laws in biology [1]. Her example of scaling laws was especially illuminating. Nevertheless, for this cherished folk-law the present author is wondering about its exceptions, too. Several examples immediately jump into his minds. One, of course, would be the folk-law itself: It had no exception in biology. This would be a bit disappointing because it would only permit us to work as what the cartographers did in J.L. Borges' fable [2]. Though eventually a map as big and as detail as the empire itself might be obtained, one would then ask where is the understanding within such an immense object? More disappointedly, such folk-law is really not biological. It has been used by some philosophers to argue against the unity of science [3].

Another example is distinctively biological and would be more exciting, and the present author believes its existence should be not surprising to biologists: *Evolution by Variation and Selection* by Darwin and Wallace. To his knowledge this dynamical law has no exception in biology. Nevertheless, under the influence of above folk-law this marvelous dynamics has been named Darwin's principle, just falling short to regard it as a fundamental law [4]. Any scientists and mathematicians would know better the difference between a principle and an equation or a law. Here the word "principle", though powerful and insightful, implies imprecision and possible fallacy, such as the Dirichlet's principle in mathematics [5]. The Darwin's principle has perhaps been used in this sense in biology [1,4]. It may be the time to change such attitude. Indeed it has been expressed that there may be general laws in biology [6], and fundamental biological laws have been articulated [7]. Based on generalizations of Fisher's fundamental theorem of natural selection [8] and Wright's adaptive landscape [9], this dynamical law of Darwin and Wallace had recently been formulated into a concise set of mathematical equations [10]. Within such formulation it had been further shown that the classical Newtonian dynamics in physics, including Newton's renowned second law, may be regarded as one of its special cases. Thus, by all accepted standards in both physical and biological sciences, the dynamical law of Darwin and Wallace has indeed achieved the status of universal laws. It will certainly play an increasingly important role in our understanding of biology at systems level. The present author would like to further speculate that such a universal law may actually be one of what physicists have been searching for during past 150 years.

For practicing biologists, the more pressing questions would be: If there are such fundamental laws, what would be their utilities in applications? Are they of any help to us to understand the diverse biological phenomena in some concrete manners? etc. Two immediate applications of the fundamental biological law may be mentioned here, which have already been elaborated from different perspectives. First, there is an issue of robustness *vs* evolvability [11-13]. It is clear that Wright's adaptive landscape provides a quantitative measure of robustness in addition to its metaphoric function. The variations encoded in the fundamental theorem of natural selection provide an instant framework to address the evolvability from two typical situations: the ability of the system to move from one adaptive peak to another and the implied possibility of stochastic change in the dynamical components. Second, there is another issue of cooperation [14-15]. The existence of multiple adaptive peaks common in evolutionary dynamics is a sufficient condition to show the ubiquitous nonlinear interaction, including cooperation. It is now well established that without cooperation it is impossible to maintain the robustness and efficiency of phage lambda genetic switch, one of most elementary biological models [16]. This genetic switch is also an example of illustrating the robustness and evolvability.

Finally, let's come back to the Borges dilemma. Borges already suggested that the one-to-one map of the imagined cartographers is not useful. The present author wishes to venture further. In his opinion the cartographer's situation would never happen in biology. We would never be able to exhaust the secrets of Nature. The number of wonders is simply too immense to be recorded down by all silicon available in our universe. Thus, high throughput and other capable tools have been developed, and will continue to be done, such as DNA sequencer [17], protein chips [18], and numerous bioinformatics platforms [19-21], to empower our ability to deeper and broader interrogate Nature and ourselves. Mega endeavors similar to Human Genome Project will be continued to be done [22-27], along with focused and small projects [16,28]. Because of the distinct mathematical characteristics in describing biological phenomena comparing to those in physical sciences, such as the explicit stochastics, combinatory, and hierarchy, in addition to the intrinsic nonlinearity, new mathematical structures is expected to be developed for better and more efficient descriptions, in addition to mathematics motivated by biology [29]. Such activities will be carried out together with our effort to understand fundamental laws in biology and with the vision enabled by such laws and by models derived from them [30]. They may be classified as in the domain of systems biology [28,31].